\documentstyle[aps]{revtex}
\begin{document}
\draft
\preprint{\vbox{Submitted to Phys.\ Rev.\ B \hfill FSU-SCRI-94-XX}}
\title{Numerical study of a Mixed Ising Ferrimagnetic System}
\author{G.~M.~Buend\'{\i}a}
\address{Supercomputer Computations Research Institute,
Florida State University, Tallahassee, Florida 32306-4052,
 and
Departamento de F\'{\i}sica, Universidad Sim\'on Bol\'{\i}var, Apdo 89000,
Caracas 1080, Venezuela\footnote{Permanent Address.}}
\author{M.~A.~Novotny}
\address{Supercomputer Computations Research Institute,
Florida State University, Tallahassee, Florida 32306-4052}
\date{\today}
\maketitle
\begin{abstract}
We present a study of a classical ferrimagnetic model on a square
lattice in which the two interpenetrating square sublattices  have
spins one-half and one.  This model
is relevant for understanding bimetallic molecular ferrimagnets that
are currently being synthesized by several experimental groups.
We perform exact ground-state calculations for the model and employ Monte
Carlo and numerical transfer-matrix techniques to obtain the
finite-temperature phase diagram for both the transition and compensation
temperatures.  When only nearest-neighbor interactions are included,
our nonperturbative results indicate no compensation point or tricritical point
at finite temperature, which contradicts earlier results obtained with
mean-field analysis.
\end{abstract}


\narrowtext


\section{Introduction}
\label{sec1}

Stable,
crystalline room-temperature magnets with spontaneous moments
are currently the subject of a great deal of interest because
of their potential device applications, such as thermomagnetic
recording\cite{re:1}.  It is widely believed that
ferrimagnetic ordering plays a fundamental role in these materials
and the synthesis of new ferrimagnets is at the moment an
active field in material science.

In a ferrimagnetic material two inequivalent moments interacting
antiferromagnetically can achieve a spontaneous
magnetization at temperatures that are low compared with the strength
of the interaction. At these low temperatures, the inequivalent
moments are antiparallel but do not cancel\cite{re:2,re:3}.
This is particularly obvious in the case of a linear chain, where
the sum of the moments in each unit cell can result in a large moment
for the chain. If adjacent chains can be positioned  such that their moments
are parallel, then a transition can occur at low temperatures to a state of
3-dimensional (3-D) ferrimagnetic order\cite{re:4}.

Important advances have been made by several groups in the synthesis of
ferrimagnetic chains\cite{re:5,re:6}.
However it is difficult to achieve high
critical temperatures with one-dimensional materials. Consequently
the discovery of
bimetallic molecular materials with spontaneous moments at temperatures as
high as 43~K \cite{re:7} has directed the experimentalist toward the formation
of 2-D and 3-D bimetallic lattices \cite{re:8}.

Synthesis of single-chain and double-chain ferrimagnets is now becoming
standard, and attempts to synthesize higher-dimensional polymeric
ferrimagnets are starting to give very encouraging results. Some of the
materials currently under investigation are
2-D organometallic ferrimagnets\cite{re:7},
2-D networks of the mixed-metal material
$\{[$P(Ph)$_4$][MnCr(ox)$_3$]\}$_n$ where Ph is phenyl and ox is
oxalate\cite{re:DSOLEGH},
3-D ferrimagnets with critical
temperatures up to 240~K\cite{re:9}, and the recently developed amorphous
$V(TCNE)_x$$\cdot$y(solvent) with ordering temperatures as high
as 400~K\cite{re:10}.

The intense activity related with the synthesis of ferrimagnetic
materials requires a parallel effort in the theoretical study of these
materials. Mixed Ising systems provide good models to study ferrimagnetism.
The magnetic properties of these model systems have been examined by
high-temperatures series expansions\cite{re:11} and
renormalization-group\cite{re:12}, mean-field\cite{re:13},  effective-field
approaches\cite{re:14a,re:14}.
An exact solution of a mixed Ising system on a union-jack lattice
(equivalent to one of the models studied in this paper) has
recently been found for a low-dimensional manifold in the
parameter space\cite{re:LIP}.
In this work we study a classical model of a
ferrimagnetic system: a mixed spin~1/2 and spin~1 system on a square lattice.
We are interested particularly in the phase diagram
and in the location and characterization of the
compensation point: the one temperature where the resultant
magnetization vanishes below the critical point (type N in the N{\'e}el
classification\cite{re:2}).  The behavior at the compensation point is of
technological significance since at this point only a small driving
field is required to change the sign of the resultant magnetization.
In a model for a ferrimagnetic thin film, it has been found that the
coercivity diverges at the compensation point\cite{re:MBU}.
Preliminary results on our model have been published elsewhere\cite{re:15}.

Experimental studies on recently synthesized compounds such as
N($n$-C$_{n}$H$_{2n+1}$)$_4$Fe$^{\rm II}$Fe$^{\rm III}$(C$_2$O$_4$)$_3$
with $n$$=$$3$--$5$ have found critical temperatures between 35 and 48~K,
and some of these compounds exhibit a compensation point near 30~K
\cite{re:MChem}.

In Sec.~\ref{sec2} we present the Hamiltonian of the model and its
ground states. We next briefly  describe the nonperturbative techniques
used to study the model: Monte Carlo
(Sec.~\ref{sec3}) and numerical transfer-matrix calculations
(Sec.~\ref{sec4}). In Sec.~\ref{sec5} we discuss our results and finally
we present the conclusions of our work in Sec.~\ref{sec6}.

\section{Model and its ground states}
\label{sec2}

Our model consists of two interpenetrating square sublattices.  One
sublattice has spins $\sigma$ on the lattice sites, where $\sigma$
has two states, $\sigma$$=$$\pm$1.  The spins $\sigma$ are spin
1/2, but we choose to put the factor of 1/2 into the interaction
parameters.
The sites of the other sublattice have spins $S$ which can have
three states, $S$$=$$\pm$1,0.  Each
spin $\sigma$ has only $S$ spins as nearest neighbors and vice versa. The
Hamiltonian includes all possible nearest and next-nearest neighbor
interactions and external fields. It is given by
\begin{eqnarray}
{\cal H}&=&
  - J_{1}\sum_{<{\rm nn}>}\sigma_{i}S_{j}
  - J_{2}\sum_{<{\rm nn}>}\sigma_{i}S_{j}^2
  - J_{3}\sum_{<{\rm nnn}>}S_{j}S_{l}\nonumber\\
&&	- J_{4}\sum_{<{\rm nnn}>}\sigma_{i}\sigma_{k}
  	- J_{5}\sum_{<{\rm nnn}>}{S_{j}^2}{S_{l}^2}
   	- \frac{J_6}{2}\sum_{<{\rm nnn}>}(S_{j}S_{l}^2
        + S_{j}^{2}S_{l})\nonumber\\
&& 	-H_{1/2}\sum_{i}\sigma_{i} - H_{1}\sum_{j}S_{j} + D\sum_{j}S_{j
}^2
\label{eqn1}
\end{eqnarray}
where the sums $\sum_{\langle{\rm nn}\rangle}$
and $\sum_{\langle{\rm nnn}\rangle}$ are over
all the nearest-neighbor (nn) and next-nearest-neighbor (nnn) bonds,
respectively.
The sums $\sum_i$ and $\sum_j$ run over all sites of the $\sigma$ and
$S$ sublattices, respectively.
Each $J$ is an exchange interaction
parameter, $D$ is the crystal field, and $H_{1}$ and $H_{1/2}$ are the
external fields, all in energy units.

{}From now on, we will label a model by enumerating the parameters
different from
zero in the Hamiltonian. For example the $J_1$-$J_2$-$D$ model is the
model with all the parameters in the above Hamiltonian zero except
$J_1$, $J_2$, and $D$.
In all cases, we fix $J_1$ to be $<$0, so the coupling
between the nn spins is antiferromagnetic.

In order to find the ground-state diagram for finite values of
the parameters, we use a 2$\times$2 cell. With
rotational symmetry taken into account, it has
$2^{2}3^{2}/2$$=$$18$ configurations.
In Table~\ref{t1} we show the 18 different configurations of the unit
cell with their respective energies and degeneracies.
Which of these states is the actual ground state depends on the values
of the parameters in the Hamiltonian.
Figs.\ \ref{fig1} through \ref{fig3} show the ground-state diagram for
different combinations of parameters. In each graph the ground-state
configurations are indicated using the notation employed in
Table~\ref{t1}. The boundaries between the regions
are obtained by pairwise equating the ground-state energies.

\section{Monte Carlo Calculations}
\label{sec3}

Standard importance sampling methods\cite{re:16} were applied to
simulate the model described by the Hamiltonian of Eq.~(\ref{eqn1})
on square lattices of $L$${\times}$$L$
sites with  periodic boundary conditions. Most of
the data were obtained with $L$$=$$40$, but we also present some results
for $L$$=$$10,16$ and $L$$=$$60$. Configurations were generated
by sequentially traversing the lattice and making single spin-flip
attempts. The flips were accepted or rejected with standard heat-bath
dynamics. We use the very long period, on the order of $2^{95}$,
random number generator KISS
(for Keep It Simple, Stupid)\cite{re:17}.  Data were generated with 25000
Monte Carlo steps per site after discarding the first 2500 steps.
The error bars were taken from the standard deviation of blocks of
500 measurements each. We define $\beta$=1/$k_{\rm B}$$T$, and take the
Boltzmann's constant $k_{\rm B}$=1. Our program calculates the internal energy
\begin{equation}
\label{eqn2}
U=\frac{1}{L^2}{\langle}{\cal H}{\rangle},
\end{equation}
the specific heat
\begin{equation}
\label{eqn3}
C=\frac{\beta^2}{L^2}[{\langle}{\cal H}^2{\rangle}-
{\langle}{\cal H}{\rangle}^2] ,
\end{equation}
the sublattice magnetizations $M_1$ and $M_2$ defined as
\begin{equation}
\label{eqn4}
M_1=\frac{2}{L^2}\left\langle\sum_{i}\sigma_i\right\rangle,
\end{equation}
\begin{equation}
\label{eqn5}
M_2=\frac{2}{L^2}\left\langle\sum_{j}S_j\right\rangle.
\end{equation}
and the total magnetization $M$$=$$1$$/$$2$($M_1$$+$$g$$M_2$), where the
factor $1$$/$$2$ gives the correct normalization for the whole lattice
since $M_1$ and $M_2$ are normalized for the sublattice.
Throughout this paper we take the $g$-factor to be $g$$=$$1$$/$$2$.
We also measured the order parameters
\begin{equation}
\label{eqn6}
O_{\pm}=\frac{1}{L^2}\left\langle\left|
\sum_{i}S_{i} \pm g\sum_{j}\sigma_{j} \right|\right\rangle
\end{equation}
and the susceptibilities associated with $M$, $M_1$, $M_2$, and $O_{\pm}$.
The averages are taken over all generated configurations,
the sums over $i$ are over all the sites with $\sigma$ spins,
and the sums over $j$ are over all the sites with $S$ spins.
There are $L^{2}/2$ terms in each sum.
We verified that our results are in agreement with exact enumeration
studies for $L$$=$$2$, and that the ground-state diagrams are reproduced for
different combinations of the parameters in the Hamiltonian.

For an infinite lattice the
order parameter $O_{+}$ would not be defined with the
absolute value in Eq.~\ref{eqn6}, and would change sign at the compensation
temperature $T_{\rm comp}$.  However, for a finite lattice the
absolute values are required to keep the order parameters nonzero in the
limit of a long measurement time.
An efficient way to locate $T_{\rm comp}$
using the Monte Carlo data is to find the crossing point between the
absolute values of the
sublattice magnetizations, i.e.,
\begin{equation}
\label{eqn7}
|M_{1}(T_{\rm comp})|=g|M_{2}(T_{\rm comp})|
\end{equation}
with the conditions
\begin{equation}
\label{eqn8}
{\rm sign}(M_{1}(T_{\rm comp}))=-{\rm sign}(M_{2}(T_{\rm comp}))
  \mbox{      ~~ and ~~     $T_{\rm comp} < T_{\rm c}$.}
\end{equation}
These relations assure that $O_{+}(T_{\rm comp})$ as defined in
Eq.~(\ref{eqn6}) is equal to zero.


\section{Transfer-Matrix Calculations}
\label{sec4}

Traditional numerical transfer-matrix (TM) calculations\cite{re:18}
were performed as a
second nonperturbative method to obtain finite-temperature
phase diagrams, critical exponents, and compensation temperatures.
These results were compared with the Monte Carlo
results, as well as with previous mean-field calculations.

For a square lattice with different spins on each of the two square
sublattices care should be taken to ensure that the TM is
symmetric.  A symmetric TM is preferred, since it is much
easier numerically to calculate its eigenvalues and eigenvectors.
We used two different
TM constructions, both of which give symmetric
transfer matrices, as detailed below.  The largest eigenvalues and
eigenvectors of these symmetric transfer matrices were then calculated
using the {\tt NAG\/} subroutine {\tt F02FJE\/}.  This subroutine requires
only multiplication of an arbitrary vector
by the TM, and consequently it is not necessary to store
the entire TM in memory.  This allows us to use
very large transfer matrices in our calculations.
For both TM implementations, the lattice is wrapped on a torus of finite
width and infinite extent, and periodic boundary conditions are
imposed.  However, the periodic boundary conditions are for a given
column of spins, and lead to different boundary conditions in terms of the
primitive lattice vectors since the torus of spins may not be along
a direction given by a single primitive lattice vector.

Two different implementations of the transfer matrix were utilized.
These are called TM1 and TM2, and the details of the construction
of the transfer matrices is given in the appendix.

The remainder of the equations in this section are stated for
the TM1 implementation with $N$, but would be equally valid for the
TM2 implementation with $N$ replaced by $\widetilde N$.
The inverse correlation length,
$\xi_{N}^{-1}$, is given by the ratio of the
largest and next-largest eigenvalues of the TM as
\begin{equation}
\label{eqn16}
\xi_N^{-1}=\ln\left|\frac{\lambda_1}{\lambda_2}\right| \; .
\end{equation}
The scaling form for $\xi$ is\cite{re:18}
\begin{equation}
\label{eqn17}
\xi_N=N{\cal F}(tN^{y_T}) ,
\end{equation}
where $t=\left|(T-T_{\rm c})/{T_{\rm c}}\right|$ is the
reduced temperature.
At $t$$=$$0$ Eq.~(\ref{eqn17}) allows one to calculate
the finite-strip estimates for the critical temperature $T_{\rm c}$
as the temperature where the phenomenological scaling relation
\begin{equation}
\label{eqn18}
\frac{\xi_N}{N}=\frac{\xi_{N+1}}{N+1}
\end{equation}
holds.
Differentiating the scaling relation Eq.~(\ref{eqn17}), and evaluating it
at the estimated value of $T_{\rm c}$ given by
Eq.~(\ref{eqn18}) gives the finite-strip estimate
of the critical exponent $y_T$$=$$1/$$\nu$ as\cite{re:18}
\begin{equation}
\label{eqn19}
y_T + 1 =
{
{\ln\left[\left.\left.d\xi_{N}/dt\right|_{t=0} \right /
\left.d\xi_{N+1}/dt\right|_{t=0}
\right]}
\over{\ln\left[N/{(N+1)}\right]}
} \; .
\end{equation}
The differentiation in Eq.~(\ref{eqn19}) was performed as a
two-point finite difference.

The TM calculation of the compensation point was only done in the
implementation TM2.
The standard method of calculating the magnetization by
diagonalizing a $2$$\times$$2$ matrix formed from the
expectation values of the magnetization operator using the two-largest
eigenvectors was used\cite{re:20}.
If ${\bf M}$ is the magnetization operator,
the solution of the equation
\begin{equation}
\label{eqnTMM}
\det\left|
\matrix{
\langle1|{\bf M}|1\rangle-m & \langle1|{\bf M}|2\rangle \hfill \cr
\langle2|{\bf M}|1\rangle \hfill & \langle2|{\bf M}|2\rangle-m \cr
}
\right| = 0
\end{equation}
give the magnetization.  Here $\langle i|$ ($|$$i$$\rangle$) is the left
(right) eigenvector associated with the $i^{th}$ largest eigenvalue of the
transfer matrix.
For a given $\widetilde N$ the
compensation temperature is the temperature below the critical temperature
where the magnetization is zero.

\section{Results}
\label{sec5}

We first tested the mean-field predictions that the $J_1$-$D$ model has a
tricritical point and a range of $D$ values with a compensation point. The
ground state for this model corresponds to the line $J_2$=0 in
Fig.\ \ref{fig1}. Figure~\ref{fig4} shows the finite-temperature
phase diagram for the $J_1$-$D$ model as a function of
$ D/|J_1|$ (remember $J_{1}$$=$$-1$) obtained with the Monte Carlo and
transfer-matrix methods, and we also show in the same graph the
mean-field results\cite{re:13}.
The transfer-matrix (TM1) results were obtained
using $N$${\times}$$\infty$ and ($N+1$)${\times}$$\infty$ lattices,
with $N$=4, 6, and 8. The Monte Carlo data for the critical temperature
were obtained from the location of the specific heat maximum for
lattices with $L$$=$$10$ and $L$$=$$16$.
These results were confirmed with $L$$=$$40$,
whose results are not shown in the figure. The finite-strip
width estimates for $y_T$ calculated with TM1 and TM2
are consistent with the Ising value $y_T$=1 and are presented in
Fig.\ \ref{fig5}. Indeed, as $N$ or ${\widetilde N}$ increases the value
for $y_T$ approaches the Ising value for all $D/|J_1|<4$.
These numerical transfer-matrix results strongly suggest that there is
a multicritical point only at $T$$=$$0$ located at the point
$D$$/$$|J_1|$$=$$4$.
There is no indication of a tricritical point at finite temperature.
It is possible that a tricritical point is located at a
much lower temperature that we could study, but even at our lowest
value of $T_{\rm c}$ we do not see any indication of a tricritical
point in the behavior of $y_T$.  An effective-field
calculation\cite{re:14a} also has found that $T_{\rm c}$$=$$0$ for
$D$$/$$|J_1|$$\ge$$4$.
Our detailed TM study shows that
$T_{\rm c}$ goes linearly to zero as
$T_{\rm c}$$/$$|J_1|$$=$$2.45$$($$\pm$$0.11$$)$$($$4$$-$$D$$/$$|J_1|$$)$ as
$D$/$|J_1|$$\rightarrow$$4$. This linear relation
holds extremely well between $D$$/$$|J_1|$$\approx$$3.8$ and our lowest value
of $D$$/$$|J_1|$$=$$3.9995$ where we get $T_{\rm c}$$/$$|J_1|$$=$$0.001225$.

A further study of the finite-temperature phase diagram for
the $J_{1}$$-$$J_{2}$$-$$D$ model leads us to the conclusion that there is
no compensation point for any range of parameters of this model.  Adding
the field interactions ($H_1$ and $H_{1/2}$) does not seem to change
this fact.  Thus we conclude that a  compensation point can not be induced
in this model by nearest-neighbor interactions.

Our next step was to include next-nearest-neighbor interactions.
Adding the $J_3$ interaction (between the $S$ spins), we found that the
finite-temperature phase diagram for the $J_1$$-$$J_3$$-$$D$ model
also showed no evidence of a compensation temperature.

Based on the above evidence, we concluded that a model with a
compensation temperature different from zero
must have, at least, interactions with non-zero parameters
$J_1$ and $J_4$ (between the $\sigma$ spins).
A general study of the finite-temperature
phase diagram of the $J_1$$-$$J_4$$-$$D$ model shows that a
compensation point exists for a certain range of
the $J_4$ parameter in the region of Fig.\ \ref{fig3}
bounded by $J_4/|J_1|>0$ and $ D/|J_1| < 4$. In Fig.\ \ref{fig6} we show
an example of the behavior of the magnetization where the
compensation and the critical point can be clearly observed.
In Fig.\ \ref{fig7} we show
the critical and compensation temperatures plotted
against $J_4$ for a particular value of $D$. The
compensation temperature does not exist until the $J_4$ interaction
takes some minimum value, after which it is almost independent of
$J_4$. This minimum value depends on $D$ as is shown in Fig.\ \ref{fig8},
where we plot as a function of $D$ the value of $J_4$ at which a compensation
point ($T_{\rm comp}<T_{\rm c}$) appears.
Fig.\ \ref{fig8} indicates that only at $D$/$J_1$$=$$4$ can there be a
compensation point without the nnn interactions ($J_4$$=$$0$). However, as
one sees in Fig.\ \ref{fig9}, the compensation temperature seems
to go to zero at this point, as does $T_{\rm c}$.

The critical temperature calculated using the numerical transfer-matrix
method (TM2) for the
$J_1$$-$$J_4$$-$$D$ model, is shown in Fig.\ \ref{fig10}(a). The Monte Carlo
estimates for $T_{\rm c}$ obtained from the maximum value of the specific
heat are in excellent agreement with the ones obtained with the transfer
matrix. To facilitate the reading of the critical temperature we present
the curves for some values of $D$ in Fig.\ \ref{fig10}(b).
For large values of $J_4$ the critical temperature seems to be independent
of $D$, but for small values of $J_4$ the dependence on $D$ is clearly
observed in the graphs. In Fig.\ \ref{fig10}(c) we show a detailed view
of the behavior of the critical temperature for a particular
value of $D$ calculated  with different transfer-matrix sizes.  The finite
size effects can be clearly appreciated. Also in this figure we
show an exact result calculated from \cite{re:LIP}. It is important
to emphasize that the exact solution is only known for the combination
of parameters shown by the dashed line in the figure. Our numerical
techniques cover the entire region of parameters.

The compensation temperature for the $J_1$$-$$J_4$$-$$D$ model is shown in
Fig.\ \ref{fig11}, where the dotted line corresponds
to Fig.\ \ref{fig9}. As before, we choose to plot the numerical
transfer-matrix results.  The Monte Carlo results are in excellent agreement.
Again the data strongly suggest that only at $D$$/$$J_1$$=$$4$ is there
a compensation temperature for $J_4$$=$$0$, and that it seems to be zero.
There is no compensation temperature at $J_4$$=$$0$ for any other value of $D$.

Small scale studies that included the other parameters of the
Hamiltonian seem to indicate that for small values of the parameters
$J_3$, $J_5$, and $J_6$ there are only minor quantitative changes
in the behavior of  $T_{\rm c}$ and $T_{\rm comp}$. However, the external
fields $H_1$ and $H_{1/2}$ play a more important role and the same seems
to happen with the parameter $J_2$. This is expected since for large
values of $D$, $J_2$ plays a role similar to an external field.

\section{Conclusions}
\label{sec6}

We have
applied two nonperturbative methods: Monte Carlo and
numerical transfer-matrix calculations to study a mixed Ising
system on a square-lattice.  The model has two interpenetrating
square sublattices, one with spins $\sigma$$=$$\pm$$1$ and the
other with spins $S$$=$$\pm$$1$$,$$0$.
The Hamiltonian has all possible fields and nn and nnn interactions.
In order to study the ferrimagnetic
behavior of the model, we choose the coupling between nearest-neighbors
to be antiferromagnetic.
We calculated exactly the ground-state phase diagrams. Also, we have
obtained the finite-temperature phase diagram and the
critical and compensation temperatures for some interesting combinations
of parameters.  We found excellent agreement between the
Monte Carlo and numerical transfer-matrix data.  Our results show
that a compensation point is induced by the presence of an interaction
$J_4$ between the spin-1/2 spins $\sigma$
(next-nearest neighbors in the lattice).
A minimum strength of the nnn interaction $J_4>0$
for a compensation point to exist was found to depend on the other
parameters of the
Hamiltonian.  We have demonstrated this in particular for the crystal field
interaction $D$.  We found that the system with only nn interactions does
not have a compensation temperature except at the point where
the crystal field takes its critical value,
$D$$/$$|J_1|$$=$$4$, and the compensation temperature and
critical temperature seem to both
be zero at this point. We also failed to find any evidence of a tricritical
point at a finite temperature. Our nonperturbative results are in
contradiction with mean-field studies for the $J_1$-$D$ model,
in which a tricritical point at finite temperatures and a range
of $D$ values with a compensation point were found\cite{re:13,re:14}.
Thus we expect that there may be regions in some
experimental two-dimensional ferrimagnets where compensation points may
vanish when the couplings between nn and nnn spins are changed,
for example by the application of external pressures.

\acknowledgments

Useful discussions with C.~P.~Landee, P.-A.~Lindg{\aa}rd, and
P.~A.~Rikvold are gratefully acknowledged.  We also thank J.~Zhang for
checking our exact ground-state calculations for
the $J_{1}$$-$$D$ model against his numerical ones.
Supported in part by the Florida State University, Supercomputer
Computations Research Institute, which is
partially funded through Contract {\#}DE-FC05-85ER25000 by the U.S.
Department of Energy, and by Florida State University through time
granted on its Cray Y-MP supercomputer.
MAN also acknowledges support from the U.S.\ D.O.E.\
grant {\#}DE-FG05-94ER45518.
GMB also acknowledges
financial support by the Venezuelan Science Counsel, CONICIT.
MAN is also supported by Department of Energy contract
{\#}DE-FG05-94ER45518 and National Science Foundation contract
{\#}DMR-9520325.

\newpage
\appendix
\section{Appendix: Construction of the Transfer Matrices}
\label{APP_TM}

This appendix presents the details of the construction of the two
implementations, labeled TM1 and TM2, of the transfer matrices that
were used in the numerical calculations.  We use the notation and
methodology of Ref.~\cite{re:19}
to show the form of the TM and that the TM is
symmetric.  It is important to realize that in this notation
curly brackets denote the matrix product introduced in
Ref.~\cite{re:19}.

The first implementation of the TM (TM1) consists of $N$ spins
$\sigma$$=$$\pm$$1$ in the first column and $N$ spins
$S$$=$$0$$,$$\pm$$1$ in the second column.  This structure is
iterated for an infinite number of columns.
This only gives a symmetric matrix only if the nnn interactions are
zero ($J_3$$=$$J_4$$=$$J_5$$=$$J_6$$=$$0$).
The TM can be written as the symmetric matrix
${\bf D}_{1/2}^{1/2}$${\bf A}$${\bf D}_1$${\bf A}^{\rm T}$${\bf D}_{1/2}^{1/2}$
where the ${\bf D}$ matrices are diagonal matrices.
The $2^N$$\times$$3^N$ matrix ${\bf A}$, written for $N$$=$$4$,
is explicitly given by
\begin{equation}
\label{eqn9}
{\bf A} = \left\{
\matrix{
{\bf Q} & {\bf Q} &         &         \cr
        & {\bf Q} & {\bf Q} &         \cr
        &         & {\bf Q} & {\bf Q} \cr
{\bf Q} &         &         & {\bf Q} \cr
}
\right\} ,
\end{equation}
where the $2$$\times$$3$ matrix ${\bf Q}$, which takes into account
the interactions with $J_1$ and $J_2$, is
\begin{equation}
\label{eqn10}
{\bf Q} = \left(
\matrix{
\exp\left[\beta( J_1+J_2)\right] & 1 & \exp\left[\beta(-J_1+J_2)\right] \cr
\exp\left[\beta(-J_1-J_2)\right] & 1 & \exp\left[\beta( J_1-J_2)\right] \cr
}
\right) \; .
\end{equation}
In the normal fashion,
the $2^N$$\times$$2^N$ diagonal matrix ${\bf D}_{1/2}$
contains interactions of the field $H_{1/2}$,
while the $3^N$$\times$$3^N$ diagonal matrix ${\bf D}_1$ contains
interactions of the fields $D$ and $H_1$.

In the second TM implementation (TM2) each column contains $\widetilde N$
spins $\sigma$ and $\widetilde N$ spins $S$.
The spins are numbered so that in one column a spin $\sigma$ is
the first spin, and in the second column a spin $S$ is the first spin.
The TM has the form
${\bf D}^{1/2}$${\bf B}$${\widetilde{\bf D}}$${\bf B}^{\rm T}$${\bf D}^{1/2}$
where each of the matrices is $6^{\widetilde N}$$\times$$6^{\widetilde N}$.
The matrix ${\bf B}$, written for $\widetilde N$$=$$3$, has the form
\begin{equation}
\label{eqn11}
{\bf B} = \left\{
\matrix{
{\bf Q} & {\bf S}         &         &         &         & {\bf S} \cr
{\bf R} & {\bf Q}^{\rm T} & {\bf R} &         &         &         \cr
        & {\bf S}         & {\bf Q} & {\bf S} &         &         \cr
        &      & {\bf R} & {\bf Q}^{\rm T} & {\bf R}    &         \cr
        &      &         & {\bf S}         & {\bf Q}    & {\bf S} \cr
{\bf R} &      &         &            & {\bf R} & {\bf Q}^{\rm T} \cr
}
\right\} ,
\end{equation}
where the $2$$\times$$3$ matrix ${\bf Q}$
is given by Eq.~(\ref{eqn10}).
The $2$$\times$$2$ matrix ${\bf S}$
takes into account nnn interactions between spins $\sigma$
(interaction $J_4$),
\begin{equation}
\label{eqn12}
{\bf S} = \left(
\matrix{
\exp\left( \beta J_4\right) & \exp\left(-\beta J_4\right) \cr
\exp\left(-\beta J_4\right) & \exp\left( \beta J_4\right) \cr
}
\right) \; ,
\end{equation}
and the $3$$\times$$3$ matrix ${\bf R}$
takes into account nnn interactions between spins $S$
(interactions $J_3$, $J_5$, and $J_6$),
\begin{equation}
\label{eqn13}
{\bf R} = \left(
\matrix{
\exp\left[\beta(J_5+J_3+J_6)\right] & 1 & \exp\left[\beta(J_5-J_3)\right] \cr
             1                   & 1 &           1                        \cr
\exp\left[\beta(J_5-J_3)\right]  & 1 & \exp\left[\beta(J_5+J_3-J_6)\right]\cr
}
\right) \; .
\end{equation}
The $6^{\widetilde N}$$\times$$6^{\widetilde N}$ diagonal matrix contains
interactions between the spins and the fields as well as nn interactions
of $J_1$ and $J_2$ within a column.  For $\widetilde N$$=$$3$ it is given by
\begin{equation}
\label{eqn14}
{\bf D} = \left\{
\matrix{
{\bf h} & {\bf Q} &        &          &         &         \cr
        & {\bf H} & {\bf Q}^{\rm T}   & &       &         \cr
        &         & {\bf h} & {\bf Q} &         &         \cr
        &         &         & {\bf H} & {\bf Q}^{\rm T} & \cr
        &         &         &         & {\bf h} & {\bf Q} \cr
{\bf Q}^{\rm T} & &         &         &         & {\bf H} \cr
}
\right\}
\end{equation}
where the $2$$\times$$2$ diagonal matrix ${\bf h}$ has diagonal
elements
$\exp(\beta H_{1/2})$ and $\exp(-\beta H_{1/2})$
and the $3\times3$ diagonal matrix ${\bf H}$ has diagonal elements
$\exp[\beta(H_1-D)]$, $1$, and $\exp[\beta(-H_1-D)]$.
The $6^{\widetilde N}$$\times$$6^{\widetilde N}$
diagonal matrix $\widetilde {\bf D}$
has the same form as Eq.~(\ref{eqn14}), but starts with ${\bf H}$ rather
than ${\bf h}$ as the first element in the matrix product.

\newpage



\figure

\begin{figure}
\caption{Ground-state diagram for the $J_1$$-$$J_2$$-$$D$ model. There are
three regions, in each of which the configurations of the $2$$\times$$2$ cells
are labeled as in Table~\protect\ref{t1}.  The transition lines not
parallel to the coordinate axes are labeled by the right-hand side of their
defining equations, $J_2$$=$$a$$D$$-$$b$$J_1$}.
\label{fig1}
\end{figure}

\begin{figure}
\caption{Ground-state diagram for the $J_1$$-$$J_3$$-$$D$ model.  The
notation is analogous to Fig.~1.}
\label{fig2}
\end{figure}

\begin{figure}
\caption{Ground-state phase diagram for the $J_1$-$J_4$-$D$ model.  The
notation is analogous to Fig.~1.}
\label{fig3}
\end{figure}

\begin{figure}
\caption{Finite-temperature phase diagram for the $J_1$$-$$D$ model. The
solid line corresponds to the mean-field
approximation.\protect\cite{re:13}
The symbols $\Diamond$ ($\ast$)
with error estimates are Monte Carlo results on lattices
with $L$$=$$10$ ($16$). The numerical transfer-matrix (TM1) results are
also shown for system size $N$$\times$$\infty$ and
($N$$+$$1$$)$$\times$$\infty$ with
$N$$=$$4$ ($\bigcirc$), $N$=6 (+), and $N$=8 ($\Box$). In the limit
$D$$\rightarrow$$-$$\infty$ the model reduces to a nn spin $\pm$$1$
 Ising model, and the arrow indicates the exact value of $T_c$ in this limit.}
\label{fig4}
\end{figure}

\begin{figure}
\caption{Numerical transfer-matrix results for the critical exponent
$y_T$$=$$1$$/$$\nu$. Finite-strip-width estimates using
$N$$\times$$\infty$
and ($N$$+$$1$)$\times$$\infty$ lattices and TM1 are
shown for $N$$=$$4$ ($\bigcirc$), $N$$=$$6$ (+), and $N$$=$$8$ ($\Box$). The
results obtained from TM2 with ${\widetilde N}$$\times$$\infty$
and (${\widetilde N}$$+$$1$$)$$\times$$\infty$ are shown for
$\widetilde N$$=$$2$ ($\times$) and $\widetilde N$$=$$3$ ($\star$).
It is clear that the results are consistent with the Ising value $y_T$$=$$1$.
This becomes more evident as we increase the size.}
\label{fig5}
\end{figure}

\begin{figure}
\caption{Magnetization vs temperature for the $J_1$$-$$J_4$$-$$D$ model
at $D$$/$$|$$J_1$$|$$=$$3.6$ and $J_4$$/$$|$$J_1$$|$$=$$3$. The
distinctive behavior of the magnetization at the compensation temperature
and at the critical temperature can be observed clearly.
The results shown are actually of $|$$M_1$$|$$-$$1$$/$$2$$|$$M_2$$|$
for system sizes $L$$=$$40$ (+) and $L$$=$$60$ ($\times$).
The ones for $M_1$$+$$1$$/$$2$$M_2$ and $O_+$ are almost identical except
close to $T_c$ where finite-size effects are largest.
The solid curves give the magnetization from Eq.~(\protect\ref{eqnTMM})
from the numerical transfer matrix calculations with
$\widetilde N$$=$$2$ through $\widetilde N$$=$$5$.
The numerical TM estimates for $T_c$ from Eq.~(\ref{eqn18}) are shown as
vertical lines for $\widetilde N$$=$$2,3$ (dashed) and $\widetilde N$$=$
$3,4$ (solid).
In the insert we  show the absolute value of the sublattice
magnetizations, $|$$M_1$$|$ and $1$$/$$2$$|$$M_2$$|$. It is clear that
at the  compensation point the two sublattice magnetization
cancel each other.  In contrast, at $T_c$ each one goes
independently to zero, except for the remanent finite-size effects.}
\label{fig6}
\end{figure}

\begin{figure}
\caption{Critical and compensation temperatures,
$T_{\rm c}$ and $T_{\rm comp}$, for the $J_1$$-$$J_4$$-$$D$
model at $D$$/$$|$$J_1$$|$$=$$3.6$.
The critical temperature is shown as a solid
line and the compensation temperature as a dotted line. Monte Carlo
results for $L$$=$$40$ are shown by the symbol $\Diamond$ with error bars.
The numerical transfer-matrix (TM2) results for $T_{\rm comp}$
 with $\widetilde N$=2 are represented by $\times$. For $T_{\rm c}$
the numerical TM sizes are $\widetilde N$$=$$2,3$ ($\times$) and
$\widetilde N$$=$$3,4$ ($\bigcirc$).
The lines are guides for the eye.}
\label{fig7}
\end{figure}

\begin{figure}
\caption{Minimum value of $J_4$ for the $J_1$$-$$J_4$$-$$D$ model
for which a compensation point exists. The symbols have the same meaning
as in Fig.\ \protect\ref{fig6}. $J_4$(minimum) seems to go to zero as
$D$$/$$|$$J_1$$|$$\rightarrow$$4$. The lines are guides to the eye.}
\label{fig8}
\end{figure}

\begin{figure}
\caption{The compensation temperature at the values of $J_4$(minimum)
given in Fig.\ \protect\ref{fig7}, shown versus $D$.
The numerical transfer-matrix data, TM2 with $\widetilde N$$=$$2$
($\times$), show that the compensation temperature approaches
zero at the point given by $J_4$$=$$0$, $D$$/$$|$$J_1$$|$$=$$4$.}
\label{fig9}
\end{figure}

\begin{figure}
\caption{(a) Critical temperature for the $J_1$$-$$J_4$$-$$D$ model.
$T_{\rm c}$ is shown as a function of $J_4$ and $D$. The values were obtained
with the numerical transfer-matrix method TM2, with $\widetilde N$=2, 3.
(b) Critical temperature for $D$$/$$|$$J_1$$|$$=$$0$,~$1$,~$2$,~$3$,~$4$
and $5$ taken from (a). The minimum of each graph depends
on the values of $J_4$ and $D$, but as
$|J_4|$ increases, $T_{\rm c}$ becomes independent of $D$.
(c) A blow-up of the $D$$/$$|$$J_1$$|$$=$$2$ region is shown.  This
illustrates the finite-size effects going from $\widetilde N$=2,3 (top
solid line) to $\widetilde N$=3,4 (lower solid line).  Also shown is
the curve (dashed line) along which the exact solution is known, and the
bicritical point ($\bigcirc$) \protect\cite{re:LIP}.
}
\label{fig10}
\end{figure}

\begin{figure}
\caption{Compensation temperature for the $J_1$$-$$J_4$$-$$D$ model,
calculated with TM2 and $\widetilde N$$=$$2$. The dotted line corresponds
to Fig.\ \protect\ref{fig9}.}
\label{fig11}
\end{figure}


\tighten
\mediumtext
 \begin{table}
\caption{Ground-state configurations, degeneracies, and energies
per lattice site for the $2$$\times$$2$ cells. The spin states are
indicated as follows: $S$$=$$0$ ($\odot$), $S$$=$$+$$1$ ($\oplus$),
$S$$=$$-$$1$ ($\ominus$), $\sigma$$=$$1$ (+) and $\sigma$$=$$-$$1$ ($-$).}

\begin{tabular}{|r|c||c|l|}
{\#} & Configuration & Degeneracy & Energy per site \\
\hline
1 & $\matrix{ \odot & + \cr + & \odot \cr}$ &
  1 & $E_1=-{1\over 2} H_{1\over 2}-J_4$\\
\hline
2 & $\matrix{ \odot & + \cr - & \odot }$ &
  2 & $E_2=J_4$\\
\hline
3 & $\matrix{ \odot & - \cr - & \odot \cr}$ &
  1 & $E_3=H_{1/2}/2 - J_4$\\
\hline
4 & $\matrix{ \oplus & + \cr + & \odot \cr}$ &
  2 & $E_4=-H_{1/2}/2 - H_{1}/4 + D/4 - J_1 - J_2 - J_4$\\
\hline
5 & $\matrix{ \oplus & + \cr - & \odot \cr}$ &
  4 & $E_5=-H_{1}/4 + D/4 + J_4$\\
\hline
6 & $\matrix{ \oplus & - \cr - & \odot \cr}$ &
  2 & $E_6=H_{1/2}/2 - H_{1}/4 + D/4 + J_1 + J_2 - J_4$\\
\hline
7 & $\matrix{ \ominus & + \cr + & \odot \cr}$ &
  2 & $E_7=-H_{1/2}/2 + H_{1}/4 + D/4 + J_1 - J_2 - J_4$\\
\hline
8 & $\matrix{ \ominus & + \cr - & \odot \cr}$ &
  4 & $E_8=H_{1}/4 + D/4 + J_4$\\
\hline
9 & $\matrix{ \ominus & - \cr - & \odot \cr}$ &
  2 & $E_9=H_{1/2}/2 + H_{1}/4 + D/4 - J_1 + J_2 - J_4$\\
\hline
10 & $\matrix{ \oplus & + \cr + & \ominus \cr}$ &
  2 & $E_{10}=-H_{1/2}/2 + D/2 - 2J_2 + J_3 - J_4 - J_5$\\
\hline
11 & $\matrix{ \oplus & + \cr - & \ominus \cr}$ &
  4 & $E_{11}=D/2 + J_3 + J_4 - J_5$\\
\hline
12 & $\matrix{ \oplus & - \cr - & \ominus \cr}$ &
  2 & $E_{12}=H_{1/2}/2 + D/2 +2J_2 + J_3 - J_4 - J_5$\\
\hline
13 & $\matrix{ \oplus & + \cr + & \oplus \cr}$ &
  1 & $E_{13}=-H_{1/2}/2 - H_{1}/2 + D/2 - 2J_1 - 2J_2
- J_3 - J_4 - J_5 - J_6$\\
\hline
14 & $\matrix{ \oplus & + \cr - & \oplus \cr}$ &
  2 & $E_{14}=-H_{1}/2 + D/2 - J_3 + J_4 - J_5 - J_6$\\
\hline
15 & $\matrix{ \oplus & - \cr - & \oplus \cr}$ &
  1 & $E_{15}=H_{1/2}/2 - H_{1}/2 + D/2 + 2J_1 + 2J_2 - J_3 - J_4 - J_5 -
J_6$\\
\hline
16 & $\matrix{ \ominus & + \cr + & \ominus \cr}$ &
  1 & $E_{16}=-H_{1/2}/2 + H_{1}/2 + D/2 + 2J_1 -2J_2 - J_3 - J_4 - J_5 +
J_6$\\
\hline
17 & $\matrix{ \ominus & - \cr + & \ominus \cr}$ &
  2 & $E_{17}=H_{1}/2 + D/2 - J_3 + J_4 - J_5 + J_6$\\
\hline
18 & $\matrix{ \ominus & - \cr - & \ominus \cr}$ &
  1 & $E_{18}=H_{1/2}/2 + H_{1}/2 + D/2 - 2J_1 + 2J_2 - J_3 - J_4 - J_5 +
J_6$\\
\end{tabular}
\label{t1}
\end{table}

\end{document}